\documentclass[ sigconf]{acmart}
\AtBeginDocument{%
  }

\usepackage{stmaryrd}
\usepackage{amsmath,amsfonts,amsbsy}

\usepackage{amssymb}
\usepackage[mathscr]{euscript}
\usepackage{hyperref}
\usepackage{url}
\usepackage{booktabs}
\usepackage{multirow}
\usepackage{float}
\usepackage[ruled,vlined,linesnumbered]{algorithm2e}
\usepackage{graphicx}
\usepackage{subfigure}
\usepackage{balance}

\newcommand{\tensor}[1]{\boldsymbol{\mathscr{#1}}}

\setcopyright{acmlicensed}
\copyrightyear{2026}
\acmYear{2026}
\acmDOI{XXXXXXX.XXXXXXX}
\acmConference[]{}{}{}

\begin{document}

\title[Jailbreaking Leaves a Trace: Understanding and Detecting Jailbreak Attacks in Large Language Models]{Jailbreaking Leaves a Trace: Understanding and Detecting Jailbreak Attacks from Internal Representations of Large Language Models}

\author{Sri Durga Sai Sowmya Kadali}
\affiliation{%
  \institution{University of California, Riverside}
  \city{Riverside}
  \state{CA}
  \country{USA}
}\email{skada009@ucr.edu}

\author{Evangelos E. Papalexakis}
\affiliation{%
  \institution{University of California, Riverside}
  \city{Riverside}
  \state{CA}
  \country{USA}
}\email{epapalex@cs.ucr.edu}


\renewcommand{\shortauthors}{Kadali et al.}

\begin{abstract}
Jailbreaking large language models (LLMs) has emerged as a critical security challenge with the widespread deployment of conversational AI systems. Adversarial users exploit these models through carefully crafted prompts to elicit restricted or unsafe outputs, a phenomenon commonly referred to as Jailbreaking. Despite numerous proposed defense mechanisms, attackers continue to develop adaptive prompting strategies, and existing models remain vulnerable. This motivates approaches that examine the internal behavior of LLMs rather than relying solely on prompt-level defenses. In this work, we study jailbreaking from both security and interpretability perspectives by analyzing how internal representations differ between jailbreak and benign prompts. We conduct a systematic layer-wise analysis across multiple open-source models, including GPT-J, LLaMA, Mistral, and the state-space model Mamba2, and identify consistent latent-space patterns associated with adversarial inputs. We then propose a tensor-based latent representation framework that captures structure in hidden activations and enables lightweight jailbreak detection without model fine-tuning or auxiliary LLM-based detectors. We further demonstrate that these latent signals can be used to actively disrupt jailbreak execution at inference time. On an abliterated LLaMA~3.1~8B model, selectively bypassing high-susceptibility layers blocks \textbf{78\%} of jailbreak attempts while preserving benign behavior on \textbf{94\%} of benign prompts. This intervention operates entirely at inference time and introduces minimal overhead, providing a scalable foundation for achieving stronger coverage by incorporating additional attack distributions or more refined susceptibility thresholds. Our results provide evidence that jailbreak behavior is rooted in identifiable internal structures and suggest a complementary, architecture-agnostic direction for improving LLM security. 
Our implementation can be found here \cite{jailbreaking_leaves_a_trace}.


\end{abstract}



\maketitle

\section{Introduction}

Large Language Models (LLMs) have demonstrated remarkable capabilities across a wide range of tasks and have become increasingly integrated into applications spanning diverse domains and user populations. Despite their utility and efforts to align them according to human and safety expectations \cite{wang2023aligninglargelanguagemodels, zeng2024autodefensemultiagentllmdefense, glaese2022improvingalignmentdialogueagents}, these models remain highly susceptible to adversarial exploitation, raising significant safety and security concerns given their widespread accessibility. Among such threats, Jailbreaking has emerged as a persistent and particularly concerning attack vector, wherein malicious actors craft carefully engineered prompts to circumvent built-in safety mechanisms \cite{chao2024jailbreakingblackboxlarge, zou2023universaltransferableadversarialattacks} and elicit restricted, sensitive, or otherwise disallowed content \cite{liu2024autodangeneratingstealthyjailbreak}. Jailbreak attacks pose substantial risks, as they enable users with harmful intent to manipulate LLMs into producing outputs that violate safety policies, including actionable instructions for malicious activities \cite{zhang2024wordgameefficienteffective, wang-etal-2024-asetf, yu2023gptfuzzer}. The growing availability of jailbreak prompts in public repositories, research artifacts, and online forums further exacerbates this issue \cite{wildteaming2024, SCBSZ24}. 

To mitigate these risks, prior work has explored a range of defense strategies, including prompt-level filtering \cite{wang2024defending}, model-level interventions \cite{zeng2024autodefensemultiagentllmdefense, xiong2025defensivepromptpatchrobust}, reinforcement learning from human feedback (RLHF)  \cite{bai2022traininghelpfulharmlessassistant}, and the use of auxiliary safety models \cite{chen2023jailbreakerjailmovingtarget, lu2024eraserjailbreakingdefenselarge}. While such approaches have demonstrated partial effectiveness, they are not without limitations. In practice, even well-aligned models can remain vulnerable under repeated or adaptive attack attempts. Moreover, no single defense mechanism has proven sufficient to counter the continually evolving landscape of jailbreak strategies. 

In this study, we investigate a complementary and comparatively underexplored direction: leveraging internal model representations to distinguish jailbreak prompts from benign inputs and to guide mitigation. Our central hypothesis is that adversarial prompts induce distinct and detectable structural patterns within the hidden representations of LLMs, independent of output behavior. To evaluate this hypothesis, we extract layer-wise internal representations such as multi-head attention and layer output/hidden state representation from multiple models such as GPT-J, LlaMa, Mistral, and the state-space sequence model Mamba, and apply tensor decomposition \cite{sidiropoulos2017tensor} techniques to characterize and compare latent-space behaviors across benign and jailbreak prompts. Building on this analysis, we further demonstrate how these latent representations can be used to identify layers that are particularly susceptible to adversarial manipulation and to intervene during inference by selectively bypassing such layers. This representation-centric framework not only enables reliable detection of jailbreak prompts but also provides a principled mechanism for mitigating harmful behavior without modifying model parameters or relying on output-level filtering. Together, our results suggest that internal representations offer a powerful and generalizable foundation for both understanding and defending against jailbreak attacks beyond surface-level text analysis. 
Our contributions are as follows:
\begin{itemize}
\item {\bf Layer-wise Jailbreak Signature}: We show that jailbreak and benign prompts exhibit distinct, layer-dependent latent signatures in the internal representations of LLMs, which can be uncovered using tensor decomposition \cite{sidiropoulos2017tensor}.
\item {\bf Effective Defense via Targeted Layer Bypass}: We demonstrate that these latent signatures can be exploited at inference time to identify susceptible layers and disrupt jailbreak execution through targeted layer bypass.
\end{itemize}

\section{Related work}

\subsection{Adversarial Attacks}
Adversarial attacks on LLMs encompass a broad class of inputs intentionally crafted to induce unintended, incorrect, or unsafe behaviors~\cite{zou2023universaltransferableadversarialattacks, chao2024jailbreakingblackboxlarge}. Unlike adversarial examples in vision or speech domains, which often rely on imperceptible input perturbations, attacks on LLMs primarily exploit semantic, syntactic, and contextual vulnerabilities in language understanding and generation. By manipulating instructions, context, or interaction structure, adversaries can steer models toward generating factually incorrect information, violating behavioral constraints, or producing harmful or sensitive content, posing significant risks to deployed systems~\cite{liu2024autodangeneratingstealthyjailbreak}. Existing adversarial strategies span a wide range of mechanisms, including prompt injection, role-playing and persona manipulation, instruction obfuscation, multi-turn coercion \cite{yu2023gptfuzzer, wildteaming2024}, and indirect attacks embedded within external content such as documents or code, and even by fine-tuning \cite{yang2023shadowalignmenteasesubverting, yao2023poisonpromptbackdoorattackpromptbased}. A central challenge in defending against these attacks is their adaptability: adversarial prompts are often transferable across models and can be easily modified to evade static defenses \cite{zou2023universaltransferableadversarialattacks}. As a result, surface-level or prompt-based mitigation strategies have shown limited robustness

\subsection{Jailbreak attacks}
A prominent and particularly challenging class of adversarial attacks on LLMs is jailbreaking. Jailbreak attacks aim to circumvent built-in safety mechanisms and alignment constraints, enabling the model to produce outputs that it is explicitly designed to refuse. These attacks often rely on prompt engineering techniques such as hypothetical scenarios, instruction overriding, contextual reframing, or step-by-step coercion, effectively manipulating the model’s internal decision-making processes \cite{wei2023jailbrokendoesllmsafety}. Unlike general adversarial prompting, jailbreak attacks explicitly target safety guardrails and content moderation policies, making them a critical concern from both security and governance perspectives \cite{liu2024autodangeneratingstealthyjailbreak, zou2023universaltransferableadversarialattacks, wildteaming2024}. Despite extensive efforts to harden models through alignment training and reinforcement learning from human feedback \cite{bai2022traininghelpfulharmlessassistant, wang2023aligninglargelanguagemodels, glaese2022improvingalignmentdialogueagents}, jailbreak prompts continue to evolve, highlighting fundamental limitations in current defense approaches. This motivates the need for methods that analyze jailbreak behavior at the level of internal model representations, rather than relying solely on external prompt or output inspection.

\subsection{Jailbreak Defenses}
Prior work on defending against jailbreak attacks in LLMs has primarily focused on prompt and output-level safeguards. Rule-based filtering and keyword matching are commonly used due to their low computational cost, but such approaches are brittle and easily bypassed through paraphrasing, obfuscation, or multi-turn prompting \cite{deng2024multilingualjailbreakchallengeslarge}. Learning-based defenses, including supervised classifiers and auxiliary LLMs for intent detection or self-evaluation \cite{wang2025selfdefendllmsdefendjailbreaking}, improve robustness but introduce additional complexity, inference overhead, and new attack surfaces. Model-level defenses, such as alignment fine-tuning, reinforcement learning from human feedback (RLHF), and policy-based or constitutional training, aim to internalize safety constraints within the model \cite{ouyang2022traininglanguagemodelsfollow}. While effective to an extent, these approaches are resource-intensive and require continual updates as jailbreak strategies evolve. Moreover, even extensively aligned models remain susceptible to jailbreak attacks, indicating fundamental limitations in current training-based defenses. Overall, existing defenses largely treat jailbreak detection as a black-box problem and rely on external signals from prompts or generated outputs. In contrast, fewer works explore the internal representations of LLMs as a basis for defense \cite{candogan2025singlepass}. This gap motivates approaches that leverage latent-space and layer-wise signals to identify jailbreak behavior in an interpretable and architecture-agnostic manner, without requiring additional fine-tuning or auxiliary models.

\section{Preliminaries}
\subsection{Tensors}
Tensors \cite{sidiropoulos2017tensor} are defined as multi-dimensional arrays that generalize one-dimensional arrays (vectors) and two-dimensional arrays (matrices) to higher dimensions. The dimension of a tensor is traditionally referred to as its \emph{order}, or equivalently, the number of \emph{modes}, while the size of each mode is called its \emph{dimensionality}. For instance, we may refer to a third-order tensor as a three-mode tensor $\tensor{X} \in \mathbb{R}^{I \times J \times K}$.

\subsection{Tensor Decomposition}
\label{sec:rank-desc}
Tensor Decomposition \cite{sidiropoulos2017tensor} is a popular data science tool for discovering underlying low-dimensional patterns in the data. We focus on the CANDECOMP/PARAFAC (CP) decomposition model \cite{sidiropoulos2017tensor}, one of the most famous tensor decomposition models that decomposes a tensor into a sum of rank-one components. We use CP decomposition because of its simplicity and interpretability. The CP decomposition of a three-mode tensor $\tensor{X} \in \mathbb{R}^{I \times J \times K}$
is the sum of three-way outer products, that is, 
$\tensor{X} \approx \sum_{r=1}^{R} \mathbf{a}_r \circ \mathbf{b}_r \circ \mathbf{c}_r$,
where $R$ is the rank of the decomposition, $\mathbf{a}_r \in \mathbb{R}^I$, $\mathbf{b}_r \in \mathbb{R}^J$, and $\mathbf{c}_r \in \mathbb{R}^K$ are the factor vectors and $\circ$ denotes the outer product. The \textit{rank} of a tensor $\tensor{X}$ is the minimal number of rank-1 tensors required to exactly reconstruct it:
\[
    \text{rank}(\tensor{X}) = \min \left\{ R : \tensor{X} = \sum_{r=1}^{R} \mathbf{a}_r \circ \mathbf{b}_r \circ \mathbf{c}_r,\; \mathbf{a}_r \in \mathbb{R}^{I}, \mathbf{b}_r \in \mathbb{R}^{J}, \mathbf{c}_r \in \mathbb{R}^{K} \right\}
    \label{eq:rank}
\]
Selecting an appropriate rank is critical, as it directly affects both the expressiveness and interpretability of the decomposition. Lower-rank approximations yield compact and computationally efficient representations, while higher ranks can capture richer structure at the cost of increased complexity and potential noise.

\subsection{Transformer Architecture}

A \textbf{Transformer} \cite{vaswani2023attentionneed} is a neural network architecture designed for modeling sequential data through attention mechanisms rather than recurrence or convolution. Transformers process input sequences in parallel and capture long-range dependencies by explicitly modeling interactions between all tokens in a sequence.

A transformer consists of a stack of layers, each composed of two primary submodules: \emph{multi-head self-attention} and a \emph{position-wise feed-forward network (FFN)}. Residual connections and layer normalization are applied around each submodule to stabilize training.

\subsubsection{Multi-Head Self-Attention}

Multi-head self-attention enables the model to attend to different parts of the input sequence simultaneously. Given an input representation $\mathbf{H} \in \mathbb{R}^{T \times d}$, each attention head projects $\mathbf{H}$ into query ($\mathbf{Q}$), key ($\mathbf{K}$), and value ($\mathbf{V}$) matrices:
\[
\mathbf{Q} = \mathbf{H}\mathbf{W}^Q, \quad
\mathbf{K} = \mathbf{H}\mathbf{W}^K, \quad
\mathbf{V} = \mathbf{H}\mathbf{W}^V.
\]
Attention is computed as
\[
\mathrm{Attn}(\mathbf{Q}, \mathbf{K}, \mathbf{V}) =
\mathrm{softmax}\!\left(\frac{\mathbf{Q}\mathbf{K}^\top}{\sqrt{d_k}}\right)\mathbf{V},
\]
where $d_k$ is the dimensionality of each attention head. Multiple attention heads operate in parallel, and their outputs are concatenated and linearly projected, allowing the model to capture diverse relational patterns across tokens.

\subsubsection{Layer Outputs and Hidden Representations}

Each transformer layer produces a \emph{hidden representation} (or layer output) that serves as input to the next layer. Formally, for layer $\ell$, the output representation $\mathbf{H}^{(\ell)}$ is given by:
\[
\mathbf{H}^{(\ell)} =
\mathrm{LN}\!\left(
\mathbf{H}^{(\ell-1)} + 
\mathrm{MHA}\!\left(\mathbf{H}^{(\ell-1)}\right)
\right),
\]
followed by
\[
\mathbf{H}^{(\ell)} =
\mathrm{LN}\!\left(
\mathbf{H}^{(\ell)} +
\mathrm{FFN}\!\left(\mathbf{H}^{(\ell)}\right)
\right),
\]
where $\mathrm{MHA}$ denotes multi-head attention, $\mathrm{FFN}$ denotes the feed-forward network, and $\mathrm{LN}$ denotes layer normalization.

The sequence of hidden states $\{\mathbf{H}^{(1)}, \ldots, \mathbf{H}^{(L)}\}$ captures increasingly abstract features, ranging from local syntactic patterns in early layers to semantic and task-relevant representations in deeper layers. These intermediate representations are commonly referred to as \emph{hidden layer activations} and form the basis for interpretability and internal behavior analysis.



\subsection{Model Types: Base, Instruction-Tuned, and Abliterated Models}

Large language models (LLMs) can be categorized based on their training and alignment processes, which influence their behavior under adversarial conditions. 

\paragraph{Base Models}  
These are pretrained on large-scale text corpora using self-supervised objectives without explicit instruction tuning. Examples include GPT-J and LLaMA. Base models capture broad language patterns but lack alignment with human preferences, making them prone to generate unrestricted or unsafe outputs.

\paragraph{Instruction-Tuned Models}  
Derived from base models via supervised fine-tuning on datasets containing human instructions and responses, these models improve instruction-following capabilities \cite{zhang2026instruction} and enforce safety constraints, such as refusing harmful queries. While instruction tuning enhances safety, these models remain susceptible to sophisticated jailbreak prompts.

\paragraph{Abliterated Models}  
Abliterated models are instruction-tuned models where alignment or safety components have been removed, disabled, or bypassed \cite{young2026comparativeanalysisllmabliteration}. Such models behave more like base models but may retain subtle differences due to prior fine-tuning. Abliterated models serve as valuable testbeds to study jailbreak vulnerabilities and internal representational changes resulting from alignment removal.

Analyzing these model types enables us to investigate how alignment and instruction tuning affect internal layer activations and latent patterns, informing the design of robust jailbreak detection methods that generalize across model variations.

\begin{figure*}[t]
    \centering
    \includegraphics[width=\textwidth]{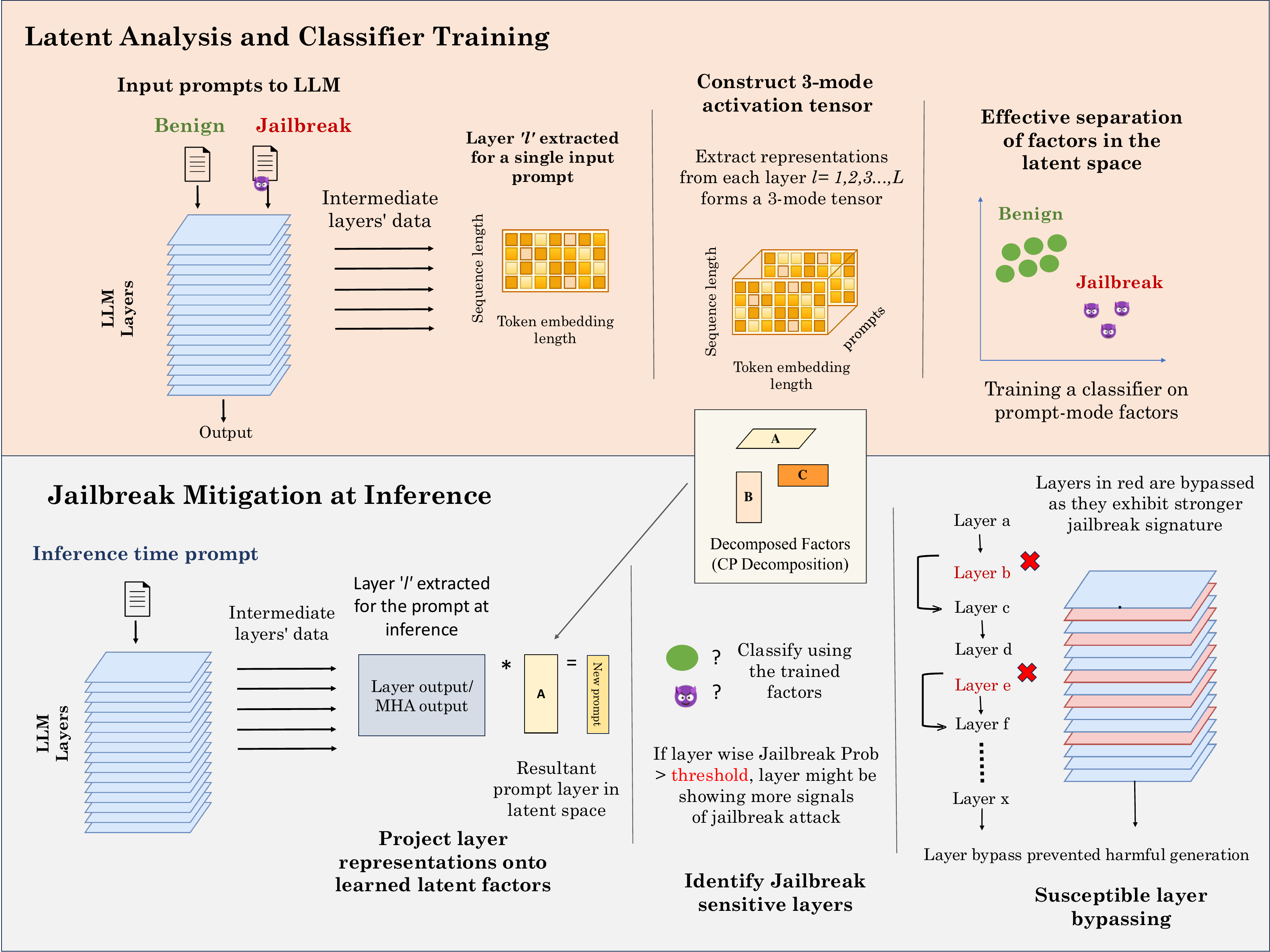}
    \caption{Proposed method: (top) Latent analysis and classifier training: self-attention and layer-output tensors are constructed from input prompts, decomposed via CP decomposition, and used to learn jailbreak-discriminative latent factors. (Bottom) Inference-time mitigation: internal representations from a new prompt are projected onto the learned factors to estimate layer-wise jailbreak susceptibility; layers exhibiting strong adversarial signals are bypassed to suppress jailbreak behavior.
    }
    \label{fig:mainfig}
\end{figure*}

\section{Proposed Method}

We study jailbreak behavior through internal model representations using two complementary analysis pipelines.


\subsection{Hidden Representation Analysis}

\paragraph{\textbf{Model Suite and Representation Extraction}}
\label{sec:part1}
We perform hidden representation analysis by examining both multi-head self-attention outputs and layer-wise hidden representations across a diverse set of large language models. Specifically, we evaluate three base models: GPT-J-6B \cite{gptj6b}, LLaMA-3.1-8B \cite{llama31}, and Mistral-7B-v01 \cite{mistral7b}; three instruction-tuned models: GPT-JT-6B \cite{gptjt6b}, LLaMA-3.1-8B-Instruct \cite{llama31instruct}, and Mistral-7B-Instruct-V0.1 \cite{mistral7binstruct}; one abliterated model, LLaMA-3.1-8B-Instruct-Abliterated \cite{llama31abliterated_hf}; and one state-space sequence model, Mamba-2.8b-hf \cite{mamba2.8b}. This selection enables a systematic comparison across different stages of alignment and architectural paradigms.

The inclusion of base, instruction-tuned, and abliterated models is intentional. Base models offer insight into unaligned latent structures; instruction-tuned models show how safety fine-tuning alters internal processing; and abliterated models help isolate the role of alignment layers. We focus not on output quality but on how jailbreak and benign prompts are internally encoded across this alignment spectrum. From this perspective, the specific semantic quality of the output is not critical; instead, we focus on identifying discriminative patterns that persist across model variants. Together, these model categories enable us to analyze jailbreak behavior across the full alignment spectrum and assess whether latent-space signatures of jailbreak prompts are consistent and model-agnostic.

\paragraph{\textbf{Tensor Construction and Latent Factors for Jailbreak Detection}}
For a given set of input prompts containing both benign and jailbreak instances, we extract multi-head attention representations and hidden state/layer output from each layer of the model. For a single prompt, the resulting representation has dimensions $1 \times T \times d$, where $T$ denotes the sequence length and $d$ is the hidden dimensionality. By stacking representations across multiple prompts, we construct a third-order tensor of size $N \times T \times d$, where $N$ is the number of prompts  as illustrated in Fig. \ref{fig:mainfig}.

To analyze the latent structure of these internal representations, we apply the CANDECOMP/PARAFAC (CP) tensor decomposition to factorize the tensor into three low-rank factors corresponding to the prompt, sequence, and hidden dimensions. The factor associated with the prompt mode captures latent patterns that reflect how different prompts are encoded internally across model layers. Prior work \cite{Papalexakis2018UnsupervisedCI, zhao-etal-2019-embedding} has shown that tensor decomposition-derived latent patterns effectively capture meaningful structure for classification and detection tasks, even with limited data \cite{ 10.1145/3589335.3651513, Kadali_2025}.

\begin{figure}[t]
    \centering
    \includegraphics[width=\linewidth]{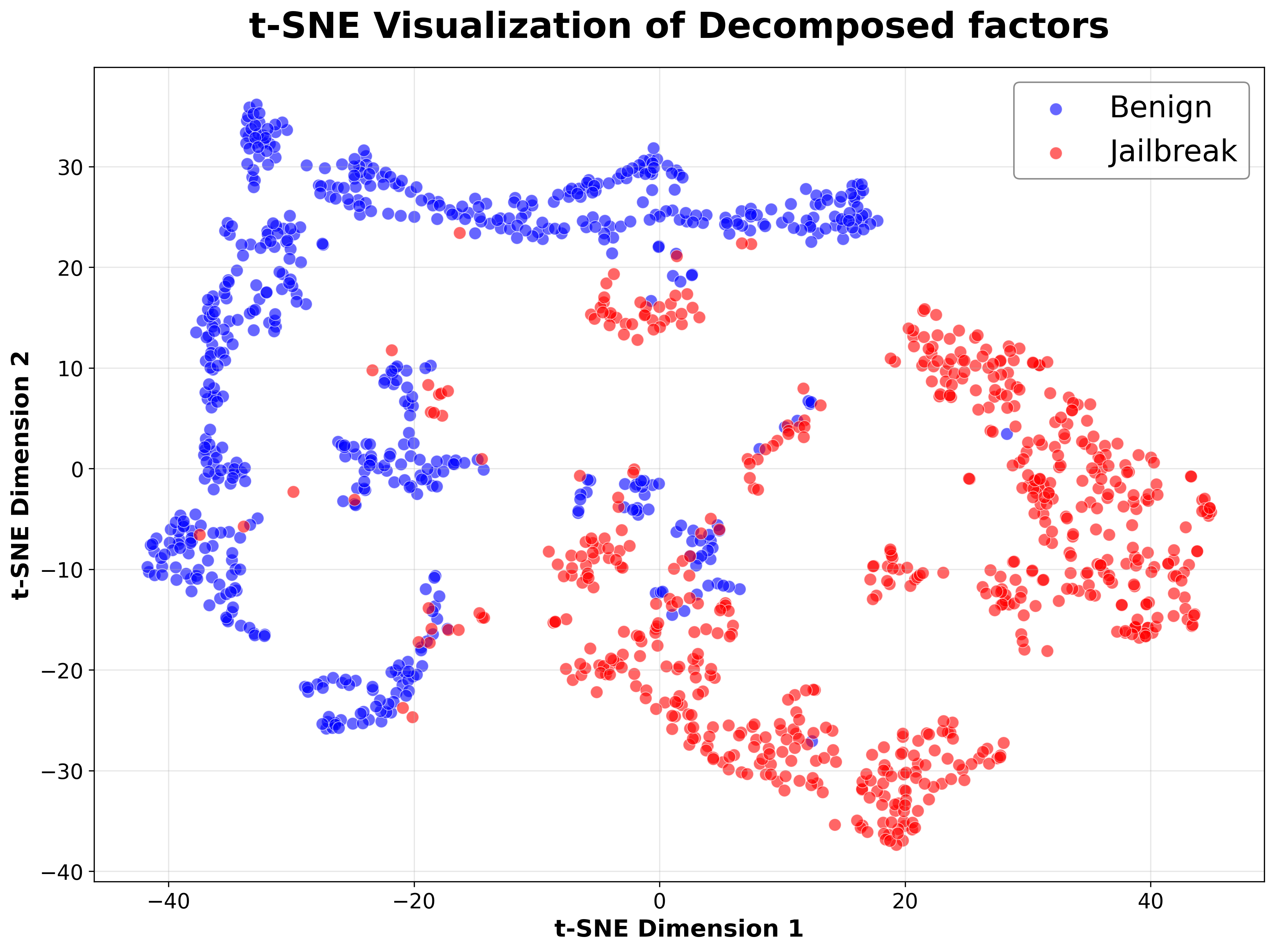}
    \caption{t-SNE visualization of prompt-mode CP factors for a representative model. Clear separation between benign and jailbreak clusters indicates that internal latent factors capture strong structure, motivating their use for jailbreak detection. Similar patterns are observed across models.
}
    \label{fig:tsne}
\end{figure}

We use these prompt-mode latent factors as features for a lightweight classifier that distinguishes jailbreak prompts from benign prompts. This classifier serves two purposes: to assess separability between prompt types in the latent space (Fig. \ref{fig:tsne}), and as a mechanism to estimate layer-wise susceptibility to jailbreak behavior, which is leveraged by the mitigation method described next.


\paragraph{\textbf{Layer-wise Separability and Susceptibility}}

To localize where jailbreak-related information is expressed within the network, we train separate classifiers on latent factors extracted from individual layers. This produces a layer-wise profile that quantifies how strongly each layer encodes representations associated with adversarial prompts. From an interpretability perspective, these layers can be viewed as critical representational stages where benign and adversarial behaviors diverge. 

Importantly, we do not interpret high separability as evidence that a layer directly causes jailbreak behavior. Rather, it indicates that these layers capture discriminative representations associated with jailbreak prompts, making them especially informative for detection. This observation provides insight into how adversarial instructions propagate through the model and forms the basis for the inference-time intervention introduced in the following section.



\subsection{Layer-Aware Mitigation via Latent-Space Susceptibility}

To mitigate jailbreak attacks, we propose a representation-level defense method, that leverages layer-wise susceptibility signals derived from internal representations. By identifying layers that strongly encode jailbreak-specific representations, we selectively bypass them during inference \cite{Elhoushi_2024, shukor2024skipping}. While prior work has explored layer bypassing primarily for reducing computation and improving inference efficiency, our approach demonstrates that such bypassing can simultaneously reduce computational cost and mitigate jailbreak behaviors \cite{luo2025adaptive, lawson2025learningskipmiddlelayers}.

We conduct this experiment on an abliterated model (LLaMA-3.1-8B). Base models are excluded from this intervention as they primarily perform next-token prediction without alignment constraints, making output-based safety evaluation less meaningful. Instruction-tuned models are also not ideal candidates, as their built-in guardrails obscure whether observed safety improvements arise from our method or from prior alignment. Abliterated models, which lack safety mechanisms while retaining instruction-tuned structure, provide a suitable testbed for isolating the effects of our approach.

\paragraph{\textbf{Layer-wise Projection and Jailbreak Susceptibility Scoring}}

Given an input prompt, we extract intermediate representations from each transformer layer of the model. Let
\(
\mathbf{x} \in \mathbb{R}^{d}
\)
denote the feature vector obtained from a specific layer (e.g., multi-head self-attention output or layer output).

\paragraph{\textbf{Latent Projection.}}
For each layer, we project the extracted features onto a lower-dimensional latent space defined by factors obtained from tensor decomposition of the corresponding instruct model representations obtained in \S\ref{sec:part1}. Let
\(
\mathbf{W} \in \mathbb{R}^{d \times r}
\)
denote the matrix of \(r\) basis vectors (factors). The projected representation is computed as:
\[
\mathbf{z} = \mathbf{W}^\top \mathbf{x},
\]
where \(\mathbf{z} \in \mathbb{R}^{r}\) is the latent feature representation. This operation constitutes a linear projection that preserves task-relevant structure encoded by the factors.
\paragraph{\textbf{Layer-wise Jailbreak Probability Estimation.}}
The projected features are passed to a classifier trained to distinguish between benign and jailbreak prompts. For a logistic regression classifier, the probability of a jailbreak at a given layer is computed as:
\[
p = \sigma(\mathbf{w}^\top \mathbf{z} + b),
\]
where \(\mathbf{w}\) and \(b\) denote the classifier weights and bias, respectively, and \(\sigma(\cdot)\) is the sigmoid function:
\[
\sigma(x) = \frac{1}{1 + e^{-x}}.
\]

\paragraph{\textbf{Classifier Training Objective.}}
The classifier is trained using labeled projected representations
\(
(\mathbf{z}_i, y_i)
\),
where \(y_i \in \{0,1\}\) indicates benign or jailbreak prompts. The parameters are optimized by minimizing the binary cross-entropy loss:
\[
\mathcal{L} = -\frac{1}{N} \sum_{i=1}^{N}
\left[
y_i \log p_i + (1 - y_i)\log(1 - p_i)
\right],
\]
where \(N\) denotes the number of samples and \(p_i\) is the predicted probability for sample \(i\).

\paragraph{\textbf{Layer Susceptibility Interpretation.}}
Layers exhibiting higher classification performance (e.g., F1 score) indicate stronger representational separability between benign and jailbreak prompts within the latent space. We interpret such layers as being more susceptible to jailbreak-style perturbations, as they encode discriminative adversarial features.

\section{Experimental Evaluation}
\label{sec:experiments}


\begin{figure*}[t]
    \centering
    \includegraphics[width=\textwidth]{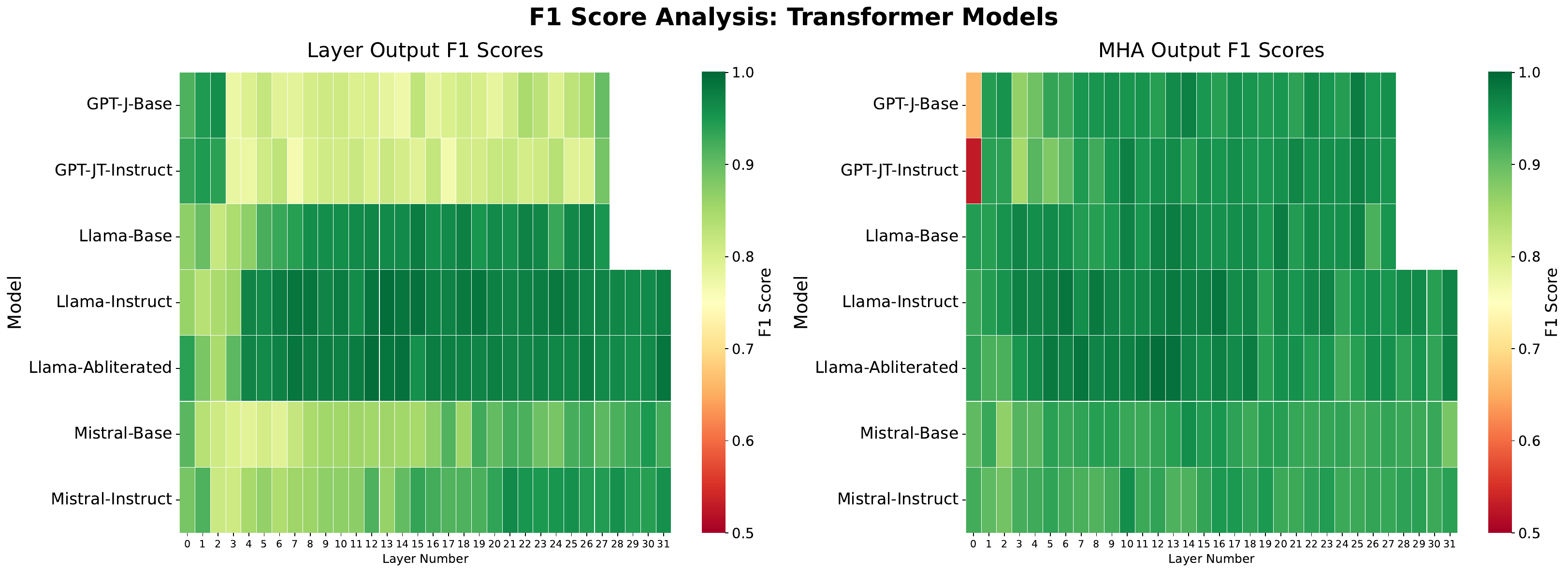}
    \caption{
(Left) Layer-wise F1 scores using CP-decomposed Transformer layer outputs.
(Right) Layer-wise F1 scores using CP-decomposed multi-head attention representations.
Jailbreak and benign prompts become reliably separable at early depths, suggesting that adversarial intent is encoded shortly after input embedding.
The strong performance of attention-based features further indicates that prompt-type information is reflected in token interaction structure as well as in hidden representations.
    }
    \label{fig:f1scores}
\end{figure*}

\subsection{Datasets and Prompt Scope}

We used two prompt sources with provenance relationships to separate representation learning from mitigation evaluation.

\textbf{Training/representation analysis.}
For latent-space analysis, we use the \texttt{Jailbreak Classification dataset} from Hugging Face \cite{jailbreakclassification_huggingface}. This dataset provides labeled benign and jailbreak prompts and is used to (i) extract layer-wise hidden representations and multi-head attention (MHA) outputs, (ii) learn CP decomposition factors for each layer and representation type, and (iii) train a lightweight classifier on the resulting latent features.

\textbf{Test/mitigation evaluation.}
To evaluate layer-aware bypass at inference time, we construct a held-out test set of 200 prompts (100 benign, 100 jailbreak) sourced from the \texttt{`In the Wild Jailbreak Prompts'} \cite{SCBSZ24} prompt collection. \cite{jailbreakclassification_huggingface} reports that its jailbreak prompts are drawn from \cite{SCBSZ24}, which motivates this choice, as these evaluations are consistent and distinct from the training corpus.

\textbf{Attack scope.}
Both datasets primarily consist of instruction-level jailbreaks (e.g., persona overrides, explicit safety negation, role-play framing, and meta-instructions). They do not include optimization-based attacks such as GCG \cite{zou2023universaltransferableadversarialattacks}, PAIR \cite{chao2024jailbreakingblackboxlarge}, or other gradient-guided adversarial suffix constructions. Since our framework operates on internal representations, extending it to additional attack families can be achieved by incorporating corresponding prompt distributions during factor learning; we leave such evaluations to future work.

\subsection{Hidden Representation Analysis}
\label{sec:analysis}
We assess whether internal model representations can reliably separate jailbreak from benign prompts across model families, layers, and representation types. Our analysis spans eight models: base, instruction-tuned, abliterated, and state-space (Mamba), providing a broad view across training paradigms.

\begin{figure}[t]
    \centering
    \includegraphics[width=\linewidth]{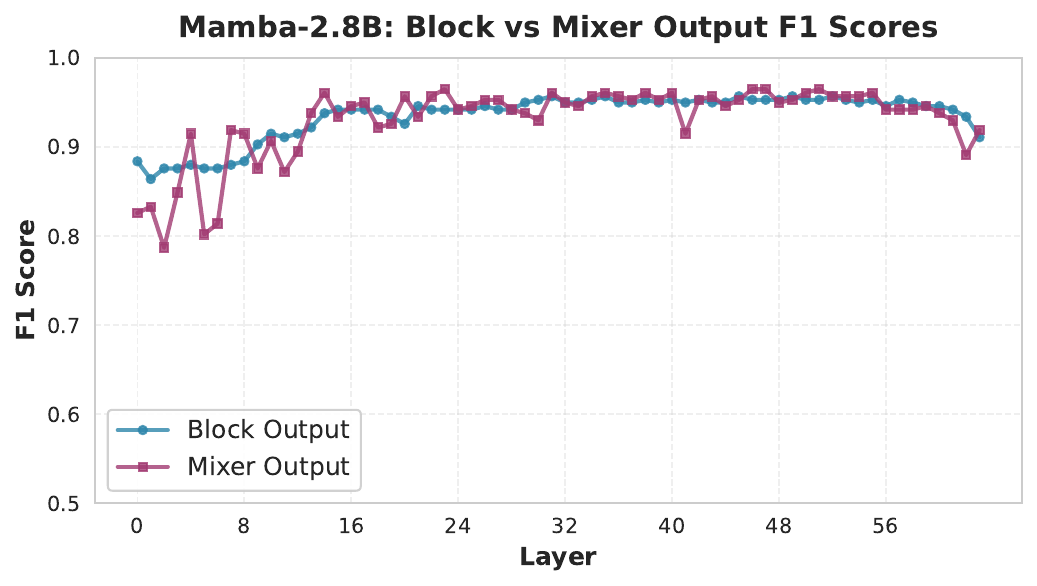}
    \caption{Layer-wise F1 scores for CP-decomposed Mamba representations (mixer and block outputs) showing early and increasing separability between benign and jailbreak prompts, indicating that state-space architectures encode adversarial prompt structure in their internal representations.
}
    \label{fig:mambaf1scores}
\end{figure}

For each model, we extract hidden states and multi-head attention (MHA) outputs across all layers. These are aggregated into third-order tensors ($N \times T \times d$) and decomposed using CP tensor decomposition (rank $r=20$) to obtain low-dimensional features for each prompt. We fix the CP decomposition rank to $r=20$ for all experiments, balancing expressiveness and efficiency as discussed in \S\ref{sec:rank-desc}, and to ensure consistent latent representations across models. A lightweight classifier is trained on these features to predict jailbreak status.  Fig.~\ref{fig:f1scores} presents layer-wise F1 scores for each model. For Mamba, we report results from both the mixer (analogous to MHA) and full block output in Fig.~\ref{fig:mambaf1scores}.

Our results show a clear separation between the two prompt types in the learned latent space, achieving consistently high F1 scores across all evaluated models. These findings suggest that jailbreak behavior manifests as identifiable and discriminative patterns within internal representations, independent of output quality or alignment stage, and can be effectively leveraged for detection without model fine-tuning.

\paragraph{\textbf{Qualitative Analysis.}}
For clarity of presentation, we visualize qualitative results for instruction-tuned models, which provide the most interpretable view of aligned internal dynamics. The qualitative patterns discussed here are representative of those observed across all evaluated models.

\begin{figure}[t]
    \centering
    \includegraphics[width=\linewidth]{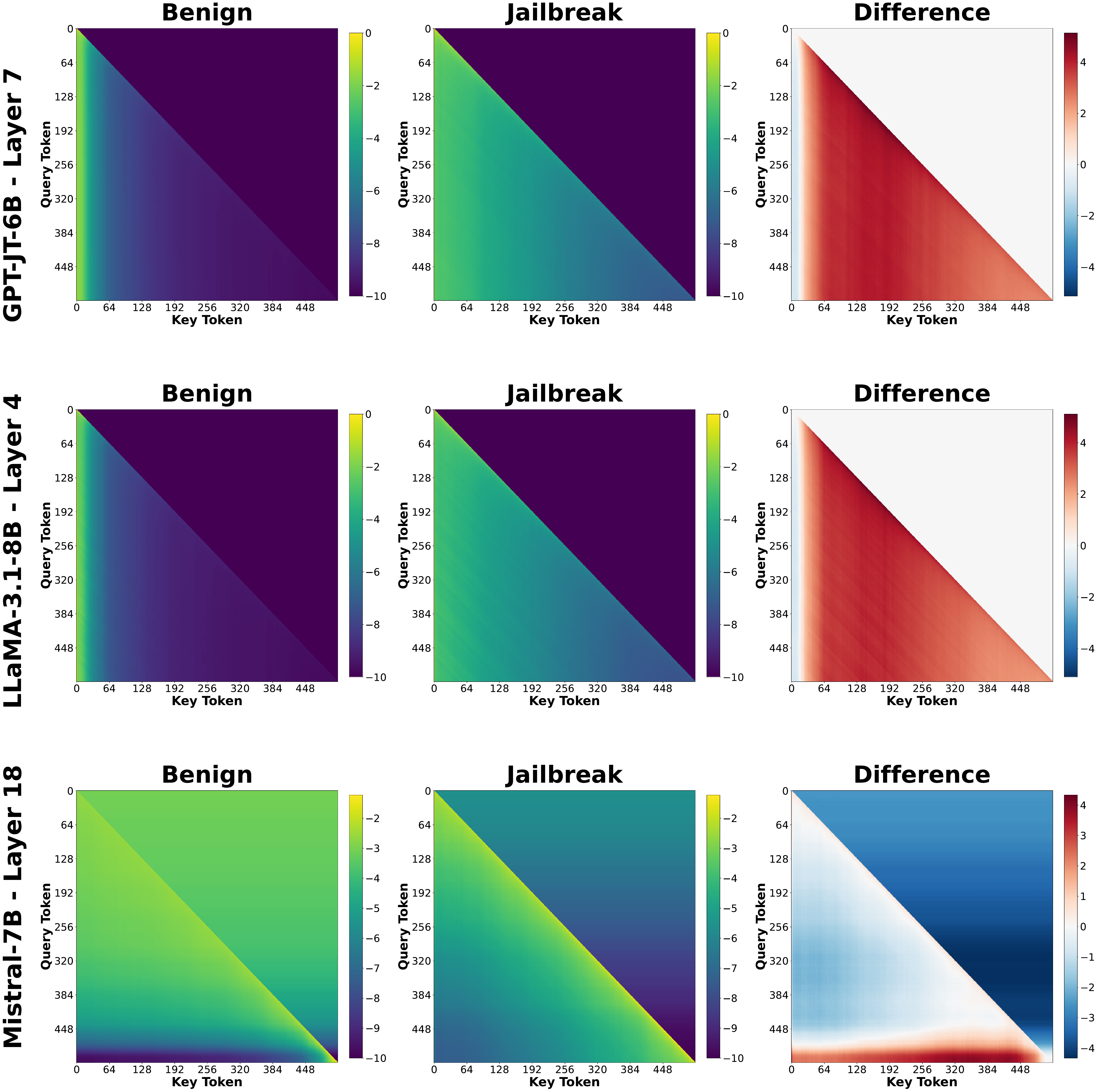}
    \caption{
Self-attention maps for three instruction-tuned models, averaged over benign and jailbreak prompts (log$_{10}$ scale).
Difference maps (right) highlight systematic but localized changes in attention patterns induced by jailbreak prompts, suggesting that adversarial intent manifests as targeted rerouting of attention rather than global disruption.
    }
    \label{fig:singlecol}
\end{figure}


\begin{figure}[t]
    \centering
    \includegraphics[width=\linewidth]{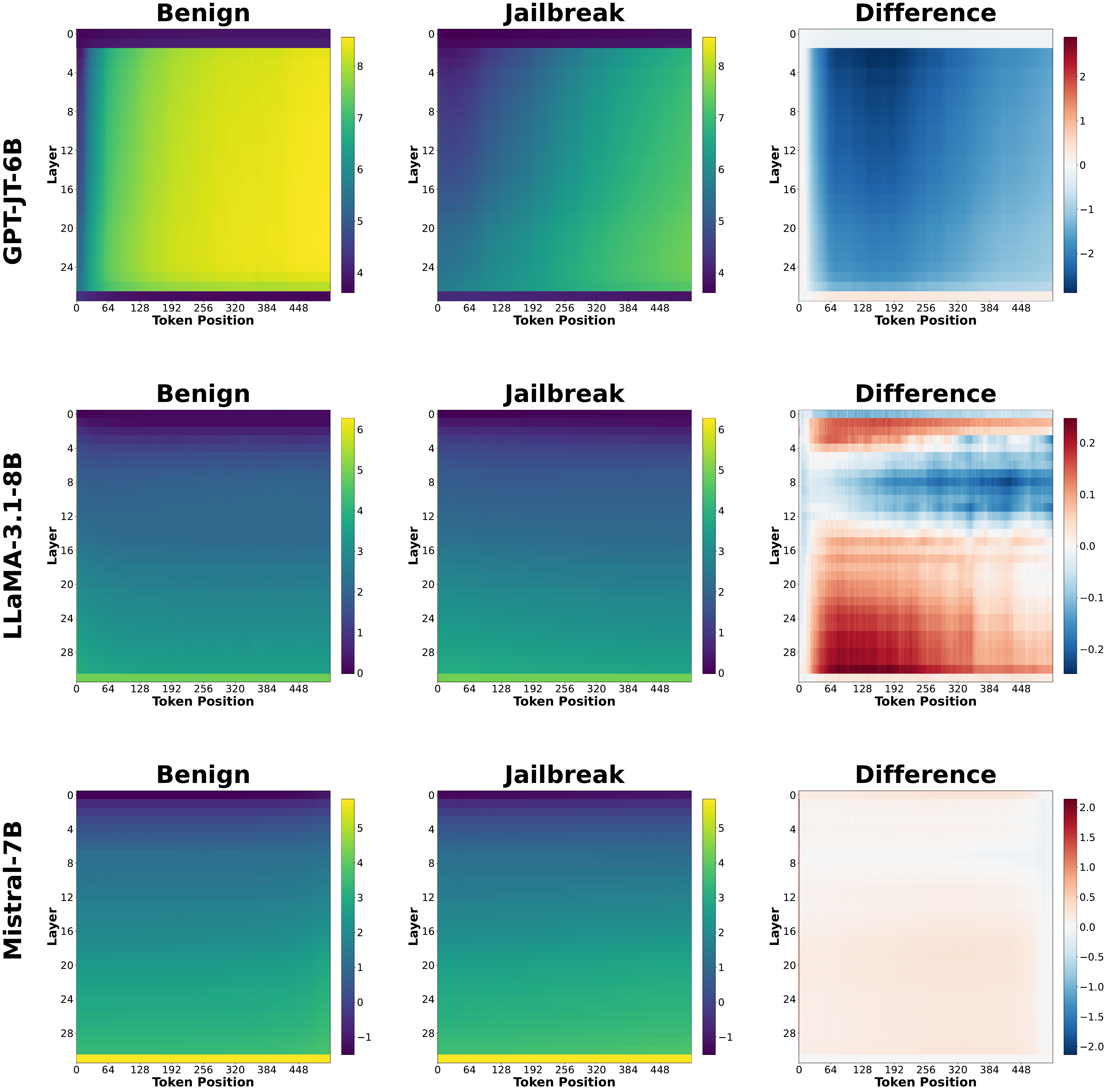}
    \caption{Layer-wise log-magnitude of hidden representations for benign (left) and jailbreak (middle) prompts, averaged across prompts, with their difference shown on the right. The difference heatmaps reveal consistent, localized deviations across layers, highlighting where adversarial prompts induce layer-dependent representational shifts.}

  
    \label{fig:attention}
\end{figure}

\paragraph{\textbf{Aggregated Self-Attention Heatmaps.}}
We use aggregated self-attention heatmaps to qualitatively assess how jailbreak prompts alter token-to-token information routing within the model. While attention alone does not encode semantic content, systematic differences in attention patterns can indicate how adversarial prompts redirect internal focus during processing.

For each instruction-tuned model and transformer layer $\ell$, we extract the self-attention weight tensors
\[
\mathbf{A}^{(\ell)} \in \mathbb{R}^{N \times H \times T \times T},
\]
where $N$ is the number of prompts, $H$ the number of attention heads, and $T$ the (padded) token length. To obtain a stable, global view of attention behavior, we aggregate over both prompts and heads.
Let $\mathcal{I}_{\text{ben}}$ and $\mathcal{I}_{\text{jb}}$ denote the index sets of benign and jailbreak prompts, respectively.
We compute the class-wise, head-averaged attention maps:
\[
\bar{\mathbf{A}}^{(\ell)}_{\text{ben}} =
\frac{1}{|\mathcal{I}_{\text{ben}}|H}
\sum_{n \in \mathcal{I}_{\text{ben}}}
\sum_{h=1}^{H}
\mathbf{A}^{(\ell)}_{n,h,:,:},
\qquad
\bar{\mathbf{A}}^{(\ell)}_{\text{jb}} =
\frac{1}{|\mathcal{I}_{\text{jb}}|H}
\sum_{n \in \mathcal{I}_{\text{jb}}}
\sum_{h=1}^{H}
\mathbf{A}^{(\ell)}_{n,h,:,:}.
\]
To highlight systematic differences between prompt types, we additionally compute a per-layer difference map:
\[
\Delta\mathbf{A}^{(\ell)} =
\bar{\mathbf{A}}^{(\ell)}_{\text{jb}} -
\bar{\mathbf{A}}^{(\ell)}_{\text{ben}}.
\]

Visualizing $\bar{\mathbf{A}}^{(\ell)}{\text{ben}}$, $\bar{\mathbf{A}}^{(\ell)}{\text{jb}}$, and $\Delta\mathbf{A}^{(\ell)}$ (Fig. \ref{fig:attention}) shows that jailbreak prompts lead to consistent, localized changes in attention patterns. This indicates that adversarial prompts influence attention by selectively emphasizing specific instruction or control tokens, providing qualitative evidence that jailbreak behavior arises from targeted changes in information flow rather than global attention disruption.


\paragraph{\textbf{Hidden-Representation Magnitude Heatmaps.}}
While attention maps reflect information routing, hidden representations capture the content and intensity of internal computation. We therefore analyze layer-wise hidden-state magnitudes to understand how strongly jailbreak prompts perturb internal activations across network depth.

For each layer $\ell$, we extract hidden states
\[
\mathbf{H}^{(\ell)} \in \mathbb{R}^{N \times T \times D},
\]
where $D$ is the hidden dimensionality.
To summarize activation strength across token positions, we compute the per-token $\ell_2$ magnitude:
\[
\mathbf{M}^{(\ell)}_{n,t} =
\left\lVert \mathbf{H}^{(\ell)}_{n,t,:} \right\rVert_{2},
\qquad
\mathbf{M}^{(\ell)} \in \mathbb{R}^{N \times T}.
\]
We then average magnitudes across prompts within each class:
\[
\bar{\mathbf{M}}^{(\ell)}_{\text{ben}}(t) =
\frac{1}{|\mathcal{I}_{\text{ben}}|}
\sum_{n \in \mathcal{I}_{\text{ben}}}
\mathbf{M}^{(\ell)}_{n,t},
\qquad
\bar{\mathbf{M}}^{(\ell)}_{\text{jb}}(t) =
\frac{1}{|\mathcal{I}_{\text{jb}}|}
\sum_{n \in \mathcal{I}_{\text{jb}}}
\mathbf{M}^{(\ell)}_{n,t}.
\]

For visualization, we apply a logarithmic transform:
\[
\tilde{\mathbf{M}}^{(\ell)}(t) =
\log\big(\bar{\mathbf{M}}^{(\ell)}(t) + \varepsilon\big),
\]
with a small $\varepsilon > 0$ for numerical stability.
We plot $\tilde{\mathbf{M}}^{(\ell)}(t)$ as heatmaps with layers on the y-axis and token positions on x-axis.

Although the averaged hidden-state magnitudes for benign and jailbreak prompts appear broadly similar (especially for LLaMA abd Mistral), their difference heatmaps reveal consistent, localized deviations across layers. This indicates that jailbreak behavior does not manifest as a global disruption of internal activations, but rather as subtle, structured changes superimposed on otherwise normal model processing.

\subsection{Layer-Aware Mitigation via Latent-Space Susceptibility}
\label{sec:exp_mitigation}

We evaluate our second proposed method, \emph{layer-aware mitigation via latent-space susceptibility}, to check whether representation-level signals can be used to suppress jailbreak execution during inference without relying on output-level filtering or fine-tuning.

\paragraph{\textbf{Experimental Setup.}}
We conduct this experiment on the \emph{abliterated} LLaMA~3.1~8B model described earlier. Evaluation is performed on a held-out set of 200 prompts (100 benign, 100 jailbreak), using latent factors learned during the analysis phase (\S\ref{sec:analysis}).

\paragraph{\textbf{Inference-time Susceptibility Scoring}}
Given an input prompt at inference time, we extract layer outputs and attention representations, project them onto the pre-learned CP factors, and use a lightweight classifier to compute a per-layer \emph{susceptibility score} indicating the strength of jailbreak-correlated features. Layers whose predicted jailbreak probability exceeds a fixed threshold ($\tau = 0.7$) are treated as \emph{highly susceptible}. The threshold $\tau$ is a tunable hyperparameter that controls the trade-off between mitigation strength and preservation of benign behavior.

\paragraph{\textbf{Layer-/Head-Bypass Intervention.}}
Based on the susceptibility score, we selectively perform: 
(\textit{i}) \textbf{Layer Bypass}: bypassing selected layer outputs; and
(\textit{ii}) \textbf{MHA Bypass}: bypassing selected attention components.
This intervention is \emph{parameter-free} (no fine-tuning), \emph{prompt-conditional} (depends on the susceptibility profile), and does not require any output-side heuristics.

\paragraph{\textbf{Output-Based Evaluation with LLM-Assisted Judging.}}
Since our goal is to prevent harmful compliance rather than optimize helpfulness, we evaluate mitigation effectiveness based on observed output behavior. Model responses are categorized as:
(i) \emph{harmful completions}, where the jailbreak intent succeeds;
(ii) \emph{benign completions}, where the model responds appropriately; and
(iii) \emph{disrupted outputs}, including truncated, repetitive, or incoherent text.

For jailbreak prompts, disrupted or non-compliant outputs are treated as successful defenses, while for benign prompts such outputs are undesirable. Output labels are assigned using an \emph{LLM-as-a-judge} rubric, followed by manual review of ambiguous cases. This evaluation protocol follows established practice for open-ended generation assessment with human validation~\cite{zheng2023judging, dubois2024lengthcontrolled, li2024llmsasjudgescomprehensivesurveyllmbased}. Based on these criteria, we define the confusion matrix as follows: 

\begin{center}
\begin{tabular}{c|c|c}
\textbf{Prompt Type} & \textbf{Observed Output} & \textbf{Outcome} \\
\hline
Jailbreak & Harmful completion & False Negative (FN) \\
Jailbreak & Disrupted/benign output & True Positive (TP) \\
Benign & Benign completion & True Negative (TN) \\
Benign & Disrupted output & False Positive (FP)
\end{tabular}
\end{center}

\paragraph{\textbf{Results.}}
Table~\ref{tab:confusion} summarizes the confusion-matrix counts. Layer-guided bypass suppresses most jailbreak attempts (\textbf{TP=78}) while largely preserving benign behavior (\textbf{TN=94}). In contrast, MHA-only bypass results in substantially more jailbreak failures (\textbf{FN=39}), indicating that layer outputs capture a larger fraction of jailbreak-relevant computation than attention components alone.

To provide a compact summary aligned with prior jailbreak evaluations, we additionally report the \emph{\textbf{attack success rate}} (\textbf{ASR}), defined as the fraction of jailbreak prompts that remain successful after mitigation. Layer-output bypass reduces ASR to $22\%$ (22/100), compared to $39\%$ for MHA-only bypass, highlighting the effectiveness of layer-level intervention.

\begin{table}[t]
\centering
\caption{Confusion matrix counts for latent-space-guided mitigation (100 jailbreak and 100 benign prompts).}
\label{tab:confusion}
\begin{tabular}{lcccc}
\toprule
Method & TP & FN & TN & FP \\
\midrule
Layer Bypass & 78 & 22 & 94 & 6 \\
MHA Bypass   & 61 & 39 & 92 & 8 \\
\bottomrule
\end{tabular}
\end{table}

\paragraph{\textbf{Failure (False Negative) Analysis}}
We examine the \textbf{22 jailbreak prompts} that remain harmful after \emph{layer-output bypass}. The majority are \textbf{persona- or roleplay-based prompt injections} (e.g., ``never refuse,'' ``no morals,'' forced speaker tags such as ``AIM:'' and ``[H4X]:'') that aim to establish persistent control over the model’s identity, tone, and formatting. Because such instructions are repeatedly reinforced throughout the prompt, elements of adversarial control can persist even when highly susceptible layers are bypassed.

Additional failures stem from \textbf{susceptibility estimation}: the intervention targets layers exceeding a fixed probability threshold chosen to preserve benign behavior. Attacks that distribute their influence across multiple layers, or weakly activate any single layer, may therefore evade suppression despite succeeding overall. Some failures also involve \textbf{milder jailbreaks} that retain adversarial framing without immediately producing explicit harmful content; under our conservative evaluation criterion, these are counted as failures.

These limitations are addressable within the proposed framework by expanding the diversity of jailbreak styles used for latent factor learning and by adopting adaptive or cumulative susceptibility criteria. Since the method operates entirely in latent space, such extensions require no architectural changes.

\section{Conclusion}

Our hypothesis and experiments indicate that internal representations of LLMs contain sufficiently strong and consistent signals to both detect jailbreak prompts and, in many cases, disrupt jailbreak execution at inference time. Importantly, these capabilities emerge from lightweight representation-level analysis and intervention, without requiring additional post-training, auxiliary models, or complex rule-based filtering. The consistency of these findings across diverse model families suggests that adversarial intent leaves stable latent-space signatures, motivating internal-representation monitoring as a practical and broadly applicable direction for understanding and mitigating jailbreak behavior.

\section{GenAI Usage Disclosure} The authors acknowledge the use of AI-based writing and coding assistance tools during the preparation of this manuscript. These tools were used exclusively to improve clarity, organization, and academic tone of text written by the authors, as well as to assist with code formatting and plot generation. All scientific ideas, methodologies, analyses, and conclusions are the original intellectual contributions of the authors. No AI system was used to generate research ideas or substantive technical content, and all AI-assisted revisions were carefully reviewed and validated by the authors.

\section{Acknowledgements} Research was supported by the National Science Foundation under CAREER grant no. IIS 2046086 and also sponsored by the Army Research Office and was accomplished under Grant Number W911NF-24-1-0397. The views and conclusions contained in this document are those of the authors and should not be interpreted as representing the official policies, either expressed or implied, of the Army Research Office or the U.S. Government. The U.S. Government is authorized to reproduce and distribute reprints for Government purposes, notwithstanding any copyright notation herein.

\bibliographystyle{ACM-Reference-Format}
\bibliography{refs}

\end{document}